\documentstyle[12pt,a4wide]{article}
\input epsf

\newcommand{\news}{\setcounter{equation}{0}}
\newcommand{\be}{\begin{equation}}
\newcommand{\ee}{\end{equation}}
\newcommand{\bea}{\begin{eqnarray}}
\newcommand{\eea}{\end{eqnarray}}
\newcommand{\bean}{\begin{eqnarray*}}
\newcommand{\eean}{\end{eqnarray*}}
\font\upright=cmu10 scaled\magstep1
\font\sans=cmss12
\newcommand{\ssf}{\sans}
\newcommand{\stroke}{\vrule height8pt width0.4pt depth-0.1pt}
\newcommand{\Z}{\hbox{\upright\rlap{\ssf Z}\kern 2.7pt {\ssf Z}}}

\newcommand{\C}{{\rlap{\rlap{C}\kern 3.8pt\stroke}\phantom{C}}}
\newcommand{\R}{\hbox{\upright\rlap{I}\kern 1.7pt R}}
\newcommand{\CP}{\C{\upright\rlap{I}\kern 1.7pt P}}
\newcommand{\half}{\frac{1}{2}}
\newcommand{\mt}{\rlap{\ssf T}\kern 3.0pt{\ssf T}}

\newcommand{\identity}{{\upright\rlap{1}\kern 2.0pt 1}}

\newcommand{\PP}{\hbox{\upright\rlap{I}\kern 1.5pt P}}
\newcommand{\an}{A_{n-1}^{(1)}}

\begin{document}
\pagestyle{plain}
\title{
\begin{flushright}
{\normalsize UKC/IMS/96-36}\\ 
{\normalsize To appear in Physics Letters B} \\
\end{flushright}
\vskip 20pt
{\Large \bf Seiberg-Witten theory, monopole spectral curves
and affine Toda solitons} \vskip 20pt}
\author{Paul M. Sutcliffe
\thanks{This work was supported in part by the 
Nuffield  Foundation}\\[20pt]
{\sl  Institute of Mathematics} \\[5pt]
{\sl University of Kent at Canterbury} \\[5pt]
{\sl  Canterbury CT2 7NZ, England} \\[20pt]
{\sl email\  P.M.Sutcliffe@ukc.ac.uk }\\[10pt]}

\date{April 1996}

\maketitle

\begin{abstract}
Using Seiberg-Witten theory 
it is known that the dynamics of $N=2$ supersymmetric
$SU(n)$ Yang-Mills theory is determined by a Riemann surface
$\Sigma_n$. In particular the mass formula for BPS states
is given by the periods of a special differential on
$\Sigma_n$.
In this note we point out that the surface $\Sigma_n$
can be obtained from the quotient of a symmetric
$n$-monopole spectral curve by its symmetry group.
Known results about the Seiberg-Witten curves then
implies that these monopoles are related to the
$A_{n-1}^{(1)}$ 
Toda lattice. We make this relation explicit via the
ADHMN construction. Furthermore, in the simplest case,
that of two $SU(2)$ monopoles, we find that the general
two monopole solution is generated by an affine Toda
soliton solution of the imaginary coupled theory.
\end{abstract}
\newpage

\section{Introduction}
\news
Seiberg and Witten \cite{SW} have determined exactly the 
nonperturbative low energy effective action
for $N=2$ supersymmetric $SU(2)$ Yang-Mills theory.
This involved the construction of a genus one surface,
or more precisely a family of such surfaces, with a
prescribed singularity behaviour. The moduli of the
surface are important, but so is the surface itself
since there exists a preferred differential such that
the spectrum of BPS states is given by integration
of this differential over the two one-cycles of the
surface.

This approach was extended \cite{KLYT} to the 
$SU(n)$ case  and involved the introduction of a
 surface $\Sigma_n$, which has genus $n-1$.
Again a special differential exists so that its
integration over the $2(n-1)$ one-cycles of
$\Sigma_n$ determines the mass formula for
BPS states. These surfaces (or equivalently the
determining curves)
were found by construction of a candidate curve
and showing that it produces the correct
quantum monodromies.

It was then observed \cite{GKMMM} that these
curves are precisely the spectral curves of
a classical integrable system, the $A_{n-1}^{(1)}$
Toda lattice. Furthermore, the special differential
emerges in the study of the Whitham dynamics
of the Toda lattice,
which essentially involves constructing a dynamics
on the moduli space of the spectral curves.

Given that the Riemann surfaces
$\Sigma_n$ are associated with a classical integrable
system, the following interesting possibility now arises.
Since we are studying a Yang-Mills theory with BPS monopoles
there is already a classical integrable system
around - namely the Bogomolny equation for static BPS
monopoles, so is it possible that the surfaces $\Sigma_n$ could
be related in some way to this integrable system ?
In the following we show that there is such a relation.
The surface $\Sigma_n$
can be obtained from the quotient of a strongly
centred $C_n$
cyclically symmetric
$n$-monopole spectral curve by its symmetry group.

The equivalence of the Toda and monopole curves thus
implies a relation between the two systems. 
We make this relation explicit, within the framework of
the ADHMN formulation, by proving that a solution of
the Toda equations determines the Nahm data of a 
cyclically symmetric monopole.

The conventional $\an$ Toda field theories do not allow
soliton solutions, but
Hollowood has shown \cite{H} that soliton solutions 
exist in the case when the coupling constant is
imaginary. We investigate in detail the simplest case
of our above correspondence, namely $SU(2)$ 2-monopoles
with $C_2$ symmetry, and find that the affine Toda
soliton solution in the imaginary coupled theory
generates precisely the known 2-monopole solution.
In fact, since every 2-monopole has 
$C_2$ symmetry about some axis, this soliton solution generates
the general solution for two monopoles. The case of
higher charge monopoles is also considered.

\section{Monopole spectral curves}
\news
We first consider classical $SU(2)$ Yang-Mills-Higgs
monopoles in the BPS limit. Hitchin has shown
\cite{Ha,Hb} that using twistor methods each static
BPS monopole with magnetic charge $n$
is equivalent to a spectral curve, defined as follows.
A spectral curve is an algebraic
curve $S \subset T\PP_1$ which has the form
\be \eta^n+\eta^{n-1} a_1(\zeta)+\ldots+\eta^r a_{n-r}(\zeta)+
\ldots+\eta a_{n-1}(\zeta)+a_n(\zeta)=0\label{algcurve}
\label{gensc}
\ee
where, for $1\leq r\leq n$, $a_r(\zeta)$ is a polynomial in $\zeta$ of
maximum degree $2r$ where $\zeta$ is the 
inhomogeneous coordinate on
the Riemann sphere and $(\zeta,\eta)$ are the standard local coordinates on
$T$\PP$_1$ defined by $(\zeta,\eta)\rightarrow\eta\frac{d}{d\zeta}$. It 
 satisfies:\\

A1. Reality condition
$$
a_r(\zeta)=(-1)^r\zeta^{2r}\overline{a_r(-\frac{1}{\overline{\zeta}})}
.$$
This is the requirement that $S$ is real with respect to the standard real structure on
$T$\PP$_1$
\be
\tau:(\zeta,\eta)\mapsto(-\frac{1}{\bar{\zeta}},
-\frac{\bar{\eta}}{\bar{\zeta}^2}).\ee

A2. $L^2$ is trivial on $S$ and $L(n-1)$ is real.\\

A3. $H^0(S,L^{\lambda}(n-2))=0$ for $\lambda\in(0,2)$.\\

Here, $L^{\lambda}$ is the holomorphic line bundle on $T$\PP$_1$ defined by the
transition function exp$(-\lambda\eta/\zeta)$ from $\zeta\not=\infty$ to
$\zeta\not=0$ and $L^{\lambda}(m)$ has transition function
 $\zeta^m\exp{(-\lambda\eta/\zeta)}$.\\

A strongly centred monopole \cite{HMM} is, roughly, one
in which the centre of mass is at the origin and
the total internal phase of the monopole is unity.
The spectral curve of a strongly centred $n$-monopole
has $a_1(\zeta)=0.$

We now consider strongly centred $n$-monopoles 
which are symmetric
under the group $G=C_n$, the cyclic group of order $n$.
The construction of spectral curves for symmetric
monopoles has been discussed in \cite{HMM}. 
The action of $G$ is generated by
\be
\eta\mapsto \omega\eta, \hskip 1cm
\zeta\mapsto \omega\zeta
\ee
where $\omega$ is an $n$th root of unity, $\omega^n=1$.
Imposing this symmetry and the reality condition on the 
spectral curve
(\ref{gensc}) results in the symmetric curve
\be
\eta^n+\eta^{n-2}\zeta^2u_2+\eta^{n-3}\zeta^3u_3
+...+\zeta^nu_n+\beta\zeta^{2n}+(-1)^n\bar\beta=0.
\label{ssc}
\ee
To satisfy the non-singularity condition A3 there
will be some relations between the real coordinates
$u_i$ and the complex coordinate $\beta$, but we shall
not be concerned with those at this stage.

What we now claim is that the quotient of this curve
by the symmetry group $G$ gives precisely the surface
$\Sigma_n$. First we show that this quotient curve
has the correct genus ie. $n-1$.
The generic $n$-monopole spectral curve is
irreducible and has genus $(n-1)^2$ \cite{Hb}.
Providing $\beta\ne 0$ the action of $G$ is free, since
there are no fixed points which lie on the curve (\ref{ssc}).
By the Riemann-Hurwitz formula the Euler characteristic
$\chi$ of the spectral curve is related to the 
 Euler characteristic $\widetilde\chi$ of the quotient
curve by
\be
\chi=\vert G\vert\widetilde\chi.
\label{rh}
\ee
So the genus $g$ of the quotient curve is determined by
\be
(2-2(n-1)^2)=n(2-2g)
\ee
giving $g=n-1$, as desired.

Now the $G$ invariants are $\eta/\zeta$ and $\zeta^n$,
so we can introduce coordinates on the quotient curve
given by 
\be
x=\eta/\zeta, \hskip 1cm z=\zeta^n\beta
\ee
to obtain the form of the quotient curve
\be
z+\mu/z+x^{n}+u_2x^{n-2}+u_3x^{n-3}+...+u_n=0
\label{todac}
\ee
where $\mu=(-1)^n\vert\beta\vert^2$.
 These are the Seiberg-Witten curves for the surfaces
$\Sigma_n$, as given in \cite{MW}.

As mentioned earlier, for the curve
(\ref{ssc}) to correspond to an $SU(2)$ BPS monopole,
that is to be a spectral curve, there are certain
non-singularity conditions to be satisfied by the moduli $u_i$.
However, these can be removed by considering
$SU(n+1)$ monopoles rather than $SU(2)$ monopoles, as follows.
For $SU(n+1)$ monopoles with maximal symmetry breaking
(ie. to $U(1)^n$)
monopoles correspond to a collection of $n$ spectral curves 
\cite{HM}. By taking a limit of this result, one
can obtain the corresponding result for the case
of minimal symmetry breaking (ie. to $U(n)$). In the special
case of $n$-monopoles it can be seen that the data in
$n-1$ of these spectral curves is trivial, so that there
is again only one spectral curve \cite{Na,HSp}. Furthermore,
the non-singularity conditions get relaxed so that the
constraining relations among the spectral curve
parameters are removed. The simplest example of this
situation, 2-monopoles in an $SU(3)$ breaking
to $U(2)$ theory, has been examined in detail by Dancer
\cite{D}.

As mentioned in the introduction it has been pointed out
\cite{GKMMM} that the curves (\ref{todac}) are the 
spectral curves of the classical $A_{n-1}^{(1)}$ Toda
lattice. Hence a relation must exist between cyclic
$n$-monopoles and the $A_{n-1}^{(1)}$ Toda lattice.
In the following section we make this relation explicit.
\section{Monopoles from affine Toda solitons}
\news
An alternative twistor formulation for monopoles
is the ADHMN construction \cite{N,Hb}, which provides
an equivalence between $n$-monopoles and Nahm data.
Here we briefly review this for the $SU(2)$ case.

Nahm data are $n\times n$ matrices 
$(T_1,T_2,T_3)$ depending
on a real parameter $s\in[0,2]$ and satisfying the following;\\

B1. Nahm's equation
\be
\frac{dT_i}{ds}=\half\sum_{j,k=1}^3\epsilon_{ijk}[T_j,T_k]. 
\label{nahm}\ee

B2. $T_i(s)$ is regular for $s\in(0,2)$ and has simple
poles at $s=0$ and $s=2$, the residues of
which form the irreducible $n$-dimensional
  representation of $su(2)$.\\

B3. $T_i(s)=-T_i^\dagger(s)$, \hskip 1cm 
 $T_i(s)=T_i^t(2-s)$.\\

Nahm's equation (\ref{nahm}) describes linear flow 
on a complex torus which is the Jacobian 
of an algebraic curve. 
In fact this algebraic curve is the monopole spectral
curve \cite{HiM} and may be explicitly 
 read off from the Nahm data as the
equation
\be
\mbox{det}(\eta+(T_1+iT_2)-2iT_3\zeta+(T_1-iT_2)\zeta^2)=0.
\label{lax}
\ee
The regularity condition B2 for $s\in(0,2)$ is the
manifestation in the ADHMN approach of the 
A3 condition for spectral curves. Again by considering
$SU(n+1)$ $n$-monopoles this condition gets
relaxed \cite{Na,HSp,D} to a regularity for $s\in(0,n+1]$
ie. the second pole in the Nahm data is lost.\\
 
It is known \cite{W} that the non-affine Toda lattice
can be obtained as a reduction of Nahm's equation.
Here we slightly extend this result to the affine case,
and moreover show that the associated monopoles have
cyclic symmetry.

To define the Toda equations \cite{OT} consider the Lie algebra
$A_r$, with $H_i$, $i=1,..,r$ the generators of the
Cartan subalgebra and $E_{\pm i}$ the generators 
corresponding to the simple roots $\alpha_i$, $i=1,..,r$.
In the Chevalley basis these satisfy 
\bea
&[H_i,E_{\pm j}]&=\pm C_{ij}E_{\pm j} \\
&[E_i,E_{-j}]&=\delta_{ij}H_j
\eea
where $C_{ij}$ are the elements of the $r\times r$
Cartan matrix for $A_r$ given by
\be
C_{ij}=\frac{2(\alpha_i,\alpha_j)}{(\alpha_i,\alpha_i)},
\hskip 1cm i,j=1,..,r.
\ee
To define the affine Toda lattice, associated with the
untwisted extended algebra $A_r^{(1)}$, we append
minus the highest root
\be
\alpha_0=-\sum_{j=1}^r\alpha_j
\ee
and its associated generator $E_{\pm 0}$ to the above
structure. We then have the same form as above, but
now indices run over the values $i=0,..,r$, also
\be
H_0=-\sum_{j=1}^rH_j
\ee
and the $r\times r$ Cartan matrix $C$ is replaced
by the $(r+1)\times (r+1)$ extended Cartan matrix $K$
with elements defined by
\be
K_{ij}=\frac{2(\alpha_i,\alpha_j)}{(\alpha_i,\alpha_i)},
\hskip 1cm i,j=0,..,r.
\ee
The connection to Nahm data is made 
by setting $r=n-1$ and introducing the ansatz
\bea
&T_1&=\frac{i\lambda}{2}\sum_{j=0}^r q_j(E_{+j}+E_{-j})
\nonumber \\
&T_2&=-\frac{\lambda}{2}\sum_{j=0}^r q_j(E_{+j}-E_{-j})
\nonumber \\
&T_3&=\frac{i\lambda}{2}\sum_{j=0}^r p_jH_j
\label{data}
\eea
where $\lambda$ is a scale parameter and the $p$'s and $q$'s
are real functions of the scaled variable $x=\lambda s$.
With this form of Nahm data, Nahm's equation (\ref{nahm})
becomes the set of equations
\be
p_i^\prime=q_i^2, \hskip 1cm 
q_i^\prime=\frac{1}{2}q_i\sum_{j=0}^r p_jK_{ij}
\label{pandq}
\ee
where prime denotes differentiation with respect to $x$.
Setting 
\be
\phi_j=2\log q_j
\ee
the above equations become the Toda lattice equation
\be
\phi_i^{\prime\prime}=\sum_{j=0}^rK_{ij}e^{\phi_j}.
\label{toda}
\ee
Taking the Nahm data (\ref{data}) and using the formula
(\ref{lax}) for the spectral curve we find that it
is exactly of the form (\ref{ssc}) for a $C_n$ symmetric
$n$-monopole.

As illustration lets consider the simplest example, that of
$n=2$. Choosing the basis
\be
H_1=\left[\begin{array}{cc}
1&0\\
0&-1
\end{array}\right]=-H_0 
\ee
the roots are
\be
\alpha_1=2=-\alpha_0
\ee
with generators $E_{+j}=E_{-j}^\dagger$ given by
\be
E_{+1}=\left[\begin{array}{cc}
0&1\\
0&0\end{array}\right]=E_{-0}. 
\ee
Thus the Nahm data has the form
$$
T_1=\frac{\lambda(q_0+q_1)}{2}\left[\begin{array}{cc}
0&i\\
i&0\end{array}\right]; \
T_2=\frac{\lambda(q_0-q_1)}{2}\left[\begin{array}{cc}
0&1\\
-1&0\end{array}\right]; 
$$
\be
T_3=-\frac{\lambda(p_0-p_1)}{2}\left[\begin{array}{ccc}
-i&0\\
0&i\end{array}\right]. 
\ee
Equation (\ref{lax}) then gives the spectral curve
\be
\eta^2+(\zeta^4+1)\lambda^2q_0q_1+\zeta^2
\lambda^2(q_0^2+q_1^2-(p_0-p_1)^2)=0.
\label{ctwo}
\ee
In fact every 2-monopole can always be translated
and rotated so that it has a spectral curve of the above form.
Of course, to be the spectral curve of an $SU(2)$ monopole
the constants appearing in (\ref{ctwo}) must take
special restricted values. We shall come to this shortly.

A couple of comments are now in order. The first is that
the Toda equation (\ref{toda}) is usually appended with
the consistent condition that $\sum_{j=0}^r \phi_j=0$,
which we also implement here. This loses no generality
due to the inclusion of the scale factor $\lambda$.
Secondly, the above version of the Toda lattice is not
quite that which is conventionally studied, due to a sign
difference. The usual case corresponds to a
hard sphere potential between the particles, in which
case there is a minus sign in front of the 
second derivative term in (\ref{toda}). Note that we
could easily convert to this situation via a Wick rotation
$x\mapsto ix$, but we shall not do this here since it
will appear more natural to handle this discrepancy
by considering equation (\ref{toda}) as the static
sector of the (1+1)-dimensional Toda field theory
\be
(\partial_t^2-\partial_x^2)\phi_i
+\sum_{j=0}^rK_{ij}e^{\phi_j}=0.
\label{todaft}
\ee
This equation does not allow soliton solutions.
Here we use the term soliton in its field theory
context as a solution which interpolates between
degenerate minima of the potential. The potential
corresponding to equation (\ref{todaft}) has
a unique minimum and hence no possibility of soliton
solutions. However, if we consider the theory with
a purely imaginary coupling constant then there are
degenerate minima of the potential and soliton solutions
indeed exist \cite{H}. Classically the coupling constant
$g$ can be absorbed into a redefinition of the field, but
we reintroduce it here to discuss the imaginary coupled
theory. Set $\phi_j=g\psi_j$ and the coupling constant $g$
appears in the equation as
\be
(\partial_x^2-\partial_t^2)\psi_i
=\frac{1}{g}\sum_{j=0}^rK_{ij}e^{g\psi_j}.
\label{todacc}
\ee
For the simplest case $n=2$, there is essentially
only one field $\psi=\psi_0=-\psi_1$ and the equation
is the sinh-Gordon equation
\be
(\partial_x^2-\partial_t^2)\psi=\frac{4}{g}\sinh g\psi
\ee
with no soliton solutions for $g$ real.
But for imaginary coupling, which we may take to be
$g=i$, the equation converts to the sine-Gordon equation
\be
(\partial_x^2-\partial_t^2)\psi=4\sin\psi
\label{sG}
\ee
which is well-known to have explicit soliton solutions.
The $n=2$ case is somewhat special in that the imaginary
coupled theory can be written as a real theory, the
sine-Gordon model. For $n>2$ the theory is intrinsically
complex (ie. a non-unitary field theory) but remarkably
the restriction to the soliton sector is unitary,
with the complex solitons having real energy \cite{H}.
Later we shall see that the reality of these affine
Toda solitons is not a problem in their relation to monopoles.

What we now show is that in the $n=2$ case the allowed
values of the constants in the spectral curve
(\ref{ctwo}) for two $SU(2)$ monopoles are determined
by the soliton solution of the affine Toda theory,
which is the sine-Gordon equation in this simplest case.
Recall we are interested in the static sector of equation
(\ref{sG}). For a 1-soliton solution the Lorentz invariance
of the equation means that a moving soliton is always
 just a boosted version of the static soliton. 
The soliton solution on the infinite line is the
well-known kink solution
\be
\psi=4\tan^{-1} e^{2x}
\label{infkink}
\ee
but here we are interested in the more general
quasi-periodic solution on a finite interval \cite{P}.
We use the results and the form of the solution as
given in \cite{Sa} 
\be
\psi=2\sin^{-1}cn(\frac{2x-\delta}{k},k)
\label{kink}
\ee
where $cn(u,k)$ denotes the Jacobi elliptic function
\cite{E}
with argument $u$ and modulus $k\in[0,1)$.
This soliton solution is quasi-periodic on an interval
of length $L$, that is $\psi(x+L)=\psi(x)+2\pi$, where
$L=kK$ and $K$ denotes the complete elliptic integral
of the first kind with modulus $k$. The constant
$\delta$ is an arbitray position coordinate. 
The infinite period solution (\ref{infkink}) is
obtained from the solution (\ref{kink}) in the limit
as $k\rightarrow 1$.

For the Nahm data generated from this soliton solution
to satisfy the condition B2, we first require that the
Nahm data is periodic in $s$ with period $2$. Since
$x=\lambda s$, and the soliton is quasi-periodic in $x$ with
period $L=kK$ this determines $\lambda$ to be $\lambda=kK/2$.
Now
\bea
&q_0=e^{\phi_0/2}&=e^{i\psi/2}=sn(\frac{2x-\delta}{k},k)
+icn(\frac{2x-\delta}{k},k) \nonumber \\
&q_1=e^{\phi_1/2}&=e^{-i\psi/2}=sn(\frac{2x-\delta}{k},k)
-icn(\frac{2x-\delta}{k},k).
\eea
Further, by equation (\ref{pandq})
\be
p_0-p_1=\partial_x\log q_0=\frac{i}{2}\partial_x\psi
\ee
hence the second constant in the spectral curve (\ref{ctwo})
has the expression
\be
q_0^2+q_1^2-(p_0-p_1)^2=2-4\sin^2(\psi/2)+\frac{1}{4}
(\partial_x\psi)^2.
\ee
But from \cite{Sa} 
\be
(\partial_x\psi)^2=16(\sin^2(\psi/2)+\frac{1-k^2}{k^2})
\ee
thus
\be
q_0^2+q_1^2-(p_0-p_1)^2=\frac{2(2-k^2)}{k^2}.
\ee
Substituting these expressions into the spectral
curve (\ref{ctwo}) we obtain the final form
\be
\eta^2+\frac{K^2}{4}(k^2(1+\zeta^4)+2(2-k^2)\zeta^2)=0
\ee
which is the known 2-monopole spectral curve \cite{AH}.
Note that we have not yet checked the reality, pole structure
and residue behaviour of our soliton generated Nahm data
but clearly since we have obtained the correct two monopole
spectral curve this must be satisfied for some choice of
the constant $\delta$. In fact taking $\delta=iK^\prime$,
where $K^\prime$ is the complete elliptic integral of the
first kind with modulus $k^\prime=\sqrt{1-k^2}$, satisfies
all the requirements and the Nahm data corresponds exactly
with a known form \cite{HSe}.

Having shown that the $C_2$ symmetric 2-monopole is
generated by the $A_1^{(1)}$ Toda soliton it is now
natural to see if a similar result holds for $n>2$.
Note that this would indeed be of interest since only 
the $n=2$ monopole spectral curves are known. Even in
 case of $n=3$ the curves are of great
interest since it is known that they include
the curve of a 3-monopole with tetrahedral
symmetry \cite{HMM}.

With a view to investigating the above issue we
consider the example of $n=3$. 
Choosing the basis
\be
H_1=\left[\begin{array}{ccc}
1&0&0\\
0&-1&0\\
0&0&0\end{array}\right]; \
H_2=\left[\begin{array}{ccc}
0&0&0\\
0&1&0\\
0&0&-1\end{array}\right]; \
H_0=\left[\begin{array}{ccc}
-1&0&0\\
0&0&0\\
0&0&1\end{array}\right] 
\ee
the extended plus simple roots are
\be
\alpha_1=(2,0), \ \alpha_2=(-1,\sqrt{3}), \ 
\alpha_0=(-1,-\sqrt{3})
\label{roots}
\ee
with generators $E_{+j}=E_{-j}^\dagger$ given by
\be
E_{+1}=\left[\begin{array}{ccc}
0&1&0\\
0&0&0\\
0&0&0\end{array}\right]; \
E_{+2}=\left[\begin{array}{ccc}
0&0&0\\
0&0&1\\
0&0&0\end{array}\right]; \
E_{+0}=\left[\begin{array}{ccc}
0&0&0\\
0&0&0\\
1&0&0\end{array}\right]. 
\ee
Thus the Nahm data has the form
$$
T_1=\frac{i\lambda}{2}\left[\begin{array}{ccc}
0&q_1&q_0\\
q_1&0&q_2\\
q_0&q_2&0\end{array}\right]; \
T_2=\frac{\lambda}{2}\left[\begin{array}{ccc}
0&-q_1&q_0\\
q_1&0&-q_2\\
-q_0&q_2&0\end{array}\right]; 
$$
\be
T_3=-\frac{i\lambda}{2}\left[\begin{array}{ccc}
p_0-p_1&0&0\\
0&p_1-p_2&0\\
0&0&p_2-p_0\end{array}\right]. 
\ee
Equation (\ref{lax}) then gives the spectral curve
\be
\eta^3+\eta\zeta^2u_2+\zeta^3u_3+\beta\zeta^6-\bar\beta=0
\ee
\bea
&\beta&=-i\lambda^3\prod_{j=0}^2q_j, \nonumber \\
&u_2&=\lambda^2\sum_{j=0}^2\ q_j^2+(p_j-p_{j+1})
(p_{j+1}-p_{j+2}) \nonumber \\
&u_3&=\lambda^3\prod_{j=0}^2(p_j-p_{j+1})
+\lambda^3\sum_{j=0}^2 q_j^2(p_{j+1}-p_{j+2})
\eea
where the indices are defined modulo 3, in accordance
with the periodicity of the lattice.

Note that the extended root system (\ref{roots})
has $C_3$ symmetry, so it is neat that it is related
to $C_3$ symmetric monopoles. Although the planar
interpretation is lost for $n>3$ the cyclic $C_n$
symmetry is still present for the extended root
system of $A_{n-1}^{(1)}$, since 
the roots have a complex representation in terms
of the $n$th roots of unity (see for eg \cite{BCDS}).
The cyclic $C_n$ symmetry is transparent in the
Dynkin diagram for  $A_{n-1}^{(1)}$, which is a closed
chain containing $n$ nodes. The example of $n=3$ is 
shown suggestively in Figure 1, to resemble three
monopoles with cyclic symmetry. It would be interesting
to consider the Toda lattice to monopoles correspondence
for other Lie algebras, and to see whether the resulting
monopoles have symmetries associated with those of the
Dynkin diagram.
Given the McKay correspondence \cite{M} it is
expected that the extended algebras of type A-D-E
are associated to monopoles with respectively
cyclic, dihedral and Platonic symmetry.

Now we have the Nahm data in terms of the Toda fields
we need to compute the static quasi-periodic soliton.
(Only solitons on the infinite line were studied
in \cite{H}). We can still work with the $\phi_i$ fields
even in the imaginary coupled theory if we regard them
as complex fields.
Defining the field $f_j=e^{\phi_j}-1$ the static Toda
field equation becomes
\be
-\partial_x^2\log(1+f_j)=f_{j+1}+f_{j-1}-2f_j.
\label{chain}
\ee
The key to constructing the soliton solution
is the remarkable elliptic function addition formula
(which can be obtained from the one given in \cite{T})
\be
dc^2(u+v)+dc^2(u-v)-2dc^2(u)=-\frac{d^2}{du^2}
\log(dc^2(u)-1-cs^2(v))
\label{ellipid}
\ee
where all the elliptic functions have the same modulus
$k$, which we suppress for clarity. 
Setting
\be
u=x+2Kj/3+\delta, \hskip 1cm v=2K/3, \hskip 1cm 
f_j=dc^2(u)-B
\ee
the identity (\ref{ellipid}) is transformed into
the equation (\ref{chain}) if the condition that
\be
B=2+cs^2(2K/3)
\ee
is also satisfied.
Hence the soliton solution is given by
\be
f_j=dc^2(x+2jK/3+\delta)-2-cs^2(2K/3)
\ee
and has period $2K$ in $x$. Note that the periodicity
of the lattice is satisfied $f_{j+3}=f_j$.
In terms of the variables for the Nahm data the solution is
\be
q_j^2=dc^2(\theta_j)-1-cs^2(2K/3), \hskip 1cm \mbox{where} 
\hskip 1cm
\theta_j=x+2jK/3+\delta.
\ee
The required periodicity in $s$ of the Nahm data can be 
achieved by a suitable choice of the scale $\lambda$,
 and the reality conditions are also satisfied. 
So the only condition that still
needs to be met is that the Nahm data has a pole at
$s=0$ whose residues define the 3-dimensional irreducible
representation of $su(2)$. It is a simple task to position
a pole at $s=0$ by choosing the constant $\delta=K$.
Then since $dc(x+K)$ has a simple pole at $x=0$
by the above formula $q_0$ has a simple pole at $s=0$.
Also by the above formulae it is clear that all the
remaining $q_i$ are regular at $s=0$. Using Taylor series 
for the elliptic functions around $x=0$ gives that
as $x\rightarrow 0$ 
\be
q_0\sim \frac{1}{x}, \ q_1\sim\frac{0}{x}, \ q_2\sim\frac{0}{x}.
\ee
Now by equation (\ref{pandq})
\be
\partial_x(p_j-p_{j+1})=q_j^2-q_{j+1}^2
\ee
which can be integrated to determine the pole behaviour
\be
p_0-p_1\sim -\frac{1}{x}, \ p_1-p_2\sim \frac{0}{x}, \
p_2-p_0\sim \frac{1}{x}.
\ee
Hence at the $s=0$ pole the behaviour of the third
Nahm matrix is $T_3\sim R_3/s$ where
\be
R_3=\frac{i}{2}\mbox{diag}(1,0,-1).
\ee
Thus $\mbox{tr}R_3^2=-1/2$ which unfortunately identifies
the representation formed by the matrix residues as
the reducible representation
 $\underline{2}\oplus\underline{1}$. So the
soliton generated Nahm matrices fall at the final hurdle
and do not correspond to Nahm data. It is easy to see
that at any pole of the soliton generated Nahm solutions
only one of the $q_i$'s can be singular, and that this
is the case for all $n$. Hence only for $n=2$ can the
representation formed by the matrix residues be irreducible.

In summary, we have shown that for all $n$ the
problem of solving Nahm's equation to determine 
cyclic $C_n$ symmetric $n$-monopoles is equivalent to solving
the quasi-periodic problem for the $A_{n-1}^{(1)}$ Toda
lattice, but that only for the case $n=2$ does the
simple soliton solution generate the Nahm data. 
For $n>2$ the Toda equation is still integrable, but
the general solution will involve abelian integrals
of genus $n-1$. The difficulty then lies in determining
those solutions which satisfy the Nahm data conditions,
which appears a highly non-trivial exercise.

\section{Conclusion}
\news
In this note we have observed that the Seiberg-Witten
curves, which are known to be equivalent to the spectral
curves of the Toda lattice, can be related to the spectral
curves of BPS monopoles. The implied connection between
Toda theory and monopoles has been made explicit 
via the ADHMN construction and in the
simplest case of two monopoles it has been shown that
the affine Toda soliton generates the known 2-monopole 
solution.

Perhaps the observations in this note are no more
than a curiosity, but there are a couple of speculative
possibilities which could be pursued. The first is that
the moduli space of a given Seiberg-Witten curve has 
singular points at which massless monopoles appear in
the supersymmetric Yang-Mills theory. It has recently
been shown \cite{HSc,Sb} that there are certain singular
points in the classical BPS moduli space where the classical
monopole solution developes additional spurious zeros
of the Higgs field. Given the identification discussed
in this note then perhaps there could be some connection
between these two phenomena.
 Secondly, through the Toda lattice connection it was
shown \cite{GKMMM} that the special Seiberg-Witten differential
emerges in the study of the Whitham dynamics
of the Toda lattice,
which essentially involves constructing a dynamics
on the moduli space of the spectral curves.
Hence is there a similar interpretation of the differential
in terms of a dynamics on the monopole moduli space,
and if so could the classical monopole moduli space metric
play a role in $N=2$ SUSY duality as it does
in S-duality in the corresponding $N=4$ SUSY theory \cite{S}.\\

\noindent{\bf Acknowledgements}

Many thanks to Nigel Hitchin for useful comments. 

\begin{figure}[ht]
\begin{center}
\leavevmode
{\epsfxsize=6cm \epsffile{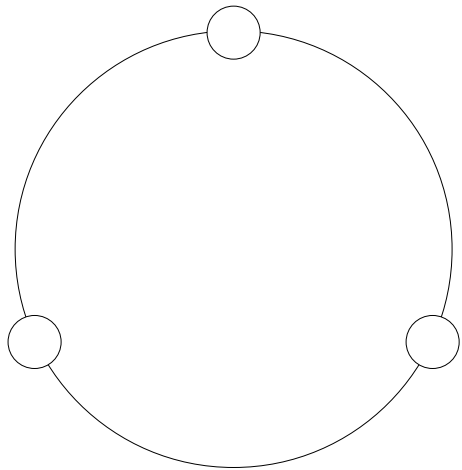}}
\end{center}
\caption{The Dynkin diagram of $A_2^{(1)}$}
\end{figure}


\newpage

\begin{thebibliography}{99}
\bibitem{AH} M.F. Atiyah and N.J. Hitchin,
\lq{\sl The geometry and dynamics of magnetic monopoles}\rq,
Princeton University Press, 1988.
\bibitem{BCDS} H.W. Braden, E. Corrigan, P.E. Dorey and
R. Sasaki, Nucl. Phys. B338, 689 (1990).
\bibitem{D} A.S. Dancer, Nonlinearity 5, 1355 (1992).
\bibitem{E} A. Erdm\a'elyi, {\sl Higher Transcendental Functions
(Bateman Project)} vol 3 (New York: McGraw-Hill).
\bibitem{GKMMM} A. Gorskii, I. Krichever, A. Marshakov,
A. Mironov and A. Morozov, Phys. Lett. B355, 466 (1995).
\bibitem{Ha} N.J. Hitchin, Commun. Math. Phys. 83, 579 (1982). 
\bibitem{Hb} N.J. Hitchin, Commun. Math. Phys. 89, 145 (1983).
\bibitem{HMM} N.J. Hitchin, N.S. Manton and M.K. Murray,
Nonlinearity, 8, 661 (1995).
\bibitem{HiM} N.J. Hitchin and M.K. Murray,
Commun. Math. Phys. 114, 463 (1988).
\bibitem{H} T.J. Hollowood, Nucl. Phys. B384, 523 (1992).
\bibitem{HSp} C.J. Houghton and P.M. Sutcliffe, in preparation.
\bibitem{HSe} C.J. Houghton and P.M. Sutcliffe, \lq{\sl
Inversion symmetric 3-monopoles and the Atiyah-Hitchin
manifold}\rq, preprint DAMTP 96-32, UKC/IMS/96-24.
\bibitem{HSc} C.J. Houghton and P.M. Sutcliffe,
 Nucl. Phys. B464, 59 (1996).
\bibitem{HM} J.Hurtubise and M.K. Murray, Commun. Math. Phys.
122, 35 (1989).
\bibitem{KLYT} A. Klemm, W. Lerche, S. Yankielowicz and
S. Theisen, Phys. Lett. B344, 169 (1995).
\bibitem{MW} E.J. Martinec and N.P. Warner, Nucl. Phys. B459, 97
(1996).
\bibitem{M} J. McKay, Proc. Sympos. Pure Math. 37, 183 (1980).
\bibitem{N} W. Nahm, \lq{\sl The construction of all self-dual
multimonopoles by the ADHM method}\rq, in Monopoles in quantum field
theory, eds. N.S. Craigie, P. Goddard and W. Nahm, World Scientific,
1982.
\bibitem{Na} H. Nakajima, \lq{\sl Monopoles and Nahm's
equation}\rq, talk presented at the British Mathematical
Colloquium, UMIST, 1996.
\bibitem{OT} D.I. Olive and N. Turok, Nucl. Phys. B220[FS8], 491
(1983).
\bibitem{P} P.D. Parmentier, in Solitons in Action, eds.
K. Lonngren and A. Scott, New York Academic, 1978.
\bibitem{SW} N. Seiberg and E. Witten, Nucl. Phys. B246, 19
(1994). Erratum-ibid B430, 485 (1994).
\bibitem{S} A. Sen, Phys. Lett. B329, 217 (1994).
\bibitem{Sa} P.M. Sutcliffe, Nonlinearity 8, 411 (1995).
\bibitem{Sb} P.M. Sutcliffe, \lq{\sl Monopole zeros}\rq, to
appear in Phys. Lett. B.
\bibitem{T} M.Toda, \lq{\sl Nonlinear waves and Solitons}\rq,
Kluwer Academic Publishers, 1989.
\bibitem{W} R.S. Ward, Phys. Lett. A112, 3 (1985).
\end{thebibliography}
\end{document}